\documentstyle[amssymb,aps,preprint]{revtex}

\begin{document}
\draft
\title{A Spatiotemporal-chaos-based Encryption Having Overall Properties
Considerably Better Than Advanced Encryption Standard}
\author{Shihong Wang$^{1,2}$, Weiping Ye$^3$, Huaping Lu$^1$, Jinyu Kuang$^3$,
Jinghua Li$^3$, and Yunlun Luo$^3$, Gang Hu$^{1,4*}$}
\address{$^1$Department of Physics, Beijing Normal University, Beijing100875, China\\
$^2$Science school, Beijing University of Posts and
Telecommunications,Beijing 100876,China\\
$^3$Department of Electronics, Beijing Normal University,
Beijing,100875,China\\
$^4$The Key Laboratory of Beam Technology and Material Modification of\\
Ministry of Education, Beijing Normal University, Beijing 100875, China\\
$^{*}$Correspondent author (Email: ganghu@bnu.edu.cn)}
\date{\today}
\maketitle

\begin{abstract}
Spatiotemporal chaos of a two-dimensional one-way coupled map lattice is
used for chaotic cryptography. The chaotic outputs of many space units are
used for encryption simultaneously. This system shows satisfactory
cryptographic properties of high security; fast encryption (decryption)
speed; and robustness against noise disturbances in communication channel.
The overall features of this spatiotemporal chaos based cryptosystem are
better than chaotic cryptosystems known so far, and also than currently used
conventional cryptosystems, such as the Advanced Encryption Standard (AES).
\end{abstract}

\pacs{PACS number(s): 05.45.Vx, 05.45.Ra, 43.72.+q}

In the last ten years of 20th century, secure communication by utilizing
chaos synchronization had attracted much attention due to the hope that the
random-like behavior of chaos and the sensitivity of chaotic trajectories to
initial conditions may provide new cryptographic methods for hiding private
information \cite{Pecora,2,3,4,6,7}. However, the recent development of
chaos communication is rather disappointing. Various cryptoanalyses have
exposed some inherent drawbacks of chaotic cryptosystems, such as low
security even with chaotic dynamics completely hidden \cite{8,9,10,11}, slow
algorithms\cite{8}, and weakness in resistance against noise disturbances in
the transmission channels, which make it difficult to promote the chaos
communication into practical service.

On the other hand, conventional cryptography has become a modern science
since the work of Shannon\cite{12}, and recently some standard and effective
cryptographic methods have been suggested, e.g., for the symmetrical scheme
Data Encryption Standard (DES) and Advanced Encryption Standard (AES issued
in 2000). In particular, the AES algorithm has been shown to have a
satisfactory trade-off in the overall properties of security, performance,
and software and hardware realizations, having no counterpart in chaotic
cryptosystems until now\cite{8,9,13}.

In our recent paper\cite{14} we have shown that a one-dimensional (1D)
coupled map lattice may reach very high practical security for chaotic
cryptography, which has in the same time fairly fast encryption and
relatively short synchronization time in comparison with other currently
known chaos-based encryption methods. However, the overall properties of the
system of\cite{14} are still not as good as AES: its encryption speed is
slower (2 to 3 times) than AES, and its avalanche effect is larger (3 to 4
times) than AES (i.e., its robustness against channel noise is weaker than
AES). In order that a chaos-based encryption method can stand up as a new
and useful cryptographic technique, it must show its significant advantages
over conventional secure communication methods, and this is the task of the
present communication.

In this paper, we propose a 2D coupled map lattice for spatiotemporal
chaotic cryptography. The most significant point is that we are able to use
a large number of chaotic sequences generated by different sites to encode
(decode) plaintexts simultaneously\cite{4}, and thus the efficiency of
cryptography can be greatly enhanced. Moreover, we can choose proper
coupling structure and encryption transformation and separate the functions
of driving bits and pure cipher bits for reaching high security together
with satisfactory robustness against channel noise. The overall properties
of our cryptosystem can be thus incomparably better than the chaotic
cryptosystems known so far, and be also considerably better than the most
effective conventional cryptographic methods, like AES.

We use the following chaotic coupled map lattice for transmitter encryption,

\begin{center}
\begin{eqnarray}
x_n(0) &=&\frac{S_n(N/2,N/2)}{2^\nu },\ \text{\ for even }N  \nonumber \\
x_{n+1}(j) &=&(1-\varepsilon )f_j[x_n(j)]+\varepsilon f_{j-1}[x_n(j-1)], 
\nonumber \\
f_0(x) &=&4x(1-x)  \nonumber \\
f_j(x) &=&a_jx(1-x),\ \ \ j=1,2,...,m  \eqnum{1a}
\end{eqnarray}

\begin{eqnarray}
z_{n+1}(0,0) &=&(1-\varepsilon )f[z_n(0,0)]+\varepsilon f_m[x_n(m)] 
\nonumber \\
y_n(0,0) &=&z_n(0,0)\times 2^h\ \ \ 
\mathop{\rm mod}%
1  \eqnum{1b} \\
f(x) &=&4x(1-x)  \nonumber
\end{eqnarray}

\begin{eqnarray}
y_{n+1}(1,0) &=&(1-\varepsilon )f[y_n(1,0)]+\varepsilon f[y_n(0,0)] 
\nonumber \\
y_{n+1}(0,1) &=&(1-\varepsilon )f[y_n(0,1)]+\varepsilon f[y_n(0,0)] 
\nonumber \\
y_{n+1}(j_1,0) &=&(1-\varepsilon )f[y_n(j_1,0)]+\varepsilon
\{0.8f[y_n(j_1-1,0)+0.2f[y_n(j_1-2,0)]\},\ \ \ j_1=2,...,N  \eqnum{1c} \\
y_{n+1}(0,j_2) &=&(1-\varepsilon )f[y_n(0,j_2)]+\varepsilon
\{0.2f[y_n(0,j_2-1)+0.8f[y_n(0,j_2-2)]\},\ \ \ j_2=2,...,N  \nonumber \\
y_{n+1}(j_1,j_2) &=&(1-\varepsilon )f[y_n(j_1,j_2)]+\varepsilon
\{0.5f[y_n(j_1-1,j_2)\ +0.5f[y_n(j_1,j_2-1)]\},\   \eqnum{1d} \\
2 &\leqslant &j_1+j_2\leqslant N  \nonumber
\end{eqnarray}
\end{center}

\begin{eqnarray}
S_n(j_1,j_2) &=&[K_n(j_1,j_2)+I_n(j_1,j_2)]\ \ \text{mod }2^\nu ,\ \  
\nonumber \\
\ K_n(j_1,j_2) &=&[\text{int}(y_n(j_1,j_2)\times 2^\mu ]\ \ \text{mod }2^\nu
,\ \ 2\leqslant j_1+j_2\leqslant N  \eqnum{1e}
\end{eqnarray}
where$\ I_n$'s are the plaintexts assumed to be private, $S_n(j_1,j_2)$ are
the ciphertexts transmitted in the open channel, and ${\bf a=(}%
a_1,a_2,...,a_m)$ are the adjustable control parameters, serving as the
secret key. For specific we will, throughout the paper, take parameters $%
h=26 $,$\ \mu =52$,$\ \nu =32$, $\ \varepsilon =0.99$, $\ m=3$, $\ N=6$, and 
$a_1=a_2=a_3=3.9$.

The schematic encoding structure is given in Fig.1. First, we use a 1D
lattice chain of length $m=3$, for setting the secret keys ${\bf a=(}%
a_1,a_2,a_3)$ in Eq.(1a). This arrangement can guarantee full sensitivity of
the encryption processes to all key parameters on one hand, and reduce as
much as possible the number of noncipher sites on the other hand. Second,
following the 1D chain is a 2D one-way coupled map lattice of Eqs.(1c) and
(1d), producing $M=25$ chaotic sequences (see the square sites of Fig.1), in
parallel for fast encryption. The 2D structure is for reducing the system
length and thus effectively reducing synchronization time and the associated
error bit avalanche. And the neighbor and next to the neighbor coupling
structure of Eq.(1c) is for enormously increasing the cost of any inverse
analytical computation attacks. Third, there are two mod operations Eqs.(1b)
and (1e) which may considerably enhance the sensitivity of chaos
synchronization to the key parameter variations, and thus reach high
practical encryption security\cite{15}. Finally, from the $M$ outputs
sequences $S_n(j_1,j_2)$, a single arbitrarily chosen sequence [$S_n(3,3)$
in Fig.1] is separated from the remaining cipher sequences as the driving
sequences in Eq.(1a) and Fig.1, and this separation will be shown later of
great significance in strengthening the robustness of the system against
channel noise. With the feed back of the driving in Eq.(1a) the encoding
system turns to be high-dimensional nonautonomous spatiotemporal chaos.

The decryption transformation of the receiver can be obtained symmetrically
by replacing $x_n(j)$, $z_n(0,0)$, $y_n(j_1,j_2)$, the key ${\bf a=(}%
a_1,a_2,...,a_m)$, with $x_n^{\prime }(j)$, $z_n^{\prime }(0,0)$, $%
y_n^{\prime }(j_1,j_2)$, and the test key ${\bf b=(}b_1,b_2,...,b_m)$,
respectively. Only one among the $M=25$ transmitted signals, $x_n^{\prime
}(0)=S_n(3,3)/2^\nu $, serves as the driving signal for the spatiotemporal
chaos synchronization of the receiver; and decoding operations become

\begin{eqnarray}
I_n^{\prime }(j_1,j_2) &=&[S_n(j_1,j_2)-K_n^{\prime }(j_1,j_2)]\ \ \text{mod 
}2^\nu ,\ \ \   \eqnum{2} \\
K_n^{\prime }(j_1,j_2) &=&[\text{int}(y_n^{\prime }(j_1,j_2)\times 2^\mu ]\
\ \text{mod }2^\nu  \nonumber
\end{eqnarray}
$I_n^{^{\prime }}$are the received plaintexts. By setting ${\bf b}={\bf a,}$
the receiver can realize synchronization to the chaotic transmitter, and
correctly extract the message as

\begin{eqnarray}
y_n^{\prime }(j_1,j_2) &=&y_n(j_1,j_2),\ \ K_n^{\prime
}(j_1,j_2)=K_n(j_1,j_2),\ \   \eqnum{3} \\
I_n^{\prime }(j_1,j_2) &=&I_n(j_1,j_2),\ 2\leqslant j_1+j_2\leqslant N 
\nonumber
\end{eqnarray}

Now, let us evaluate our cryptosystem in the aspects of performance,
security and robustness. The first significant point of Eqs.(1) and Eq.(2)
is that we fully take the advantage of spatiotemporal chaos in the
performance. A large number of chaotic sites in the lattice network can be
used for encryption (decryption) in parallel. For the present parameters, we
have totally 31 coupled maps, among which 25 chaotic sites can produce
ciphertexts simultaneously. Thus, the encryption (decryption) efficiency is
extremely high. Specially, we can produce 350-Mbit ciphers per second with
our 750MHz CPU computer. In comparison, for conventional block ciphers, AES
can have encryption speed of 96-Mbit, 80-Mbit, and 66-Mbit ciphers per
second for 128-bit, 192-bit, and 256-bit key sizes (for a 600 MHz CPU PC) 
\cite{13}, respectively, and for conventional stream ciphers a 32-bit linear
feedback shift register (LFSR), which has very low security, has encryption
speed of 20-Mbits per second with our same PC.

A crucial problem for the scheme of Fig.1 is whether $M=25$ keystreams can
be effectively used in parallel. A positive answer can be available only if
different keystreams are practically uncorrelated. In Figs.2(a) and 2(b) we
plot the mutual correlation $C_{24,23}(\tau )$ (for the definition, see
Eq.(3.2) of \cite{4}) and mutual information $%
I(K_n(2,4);K_n(2,3))=H(K_n(2,4))-H(K_n(2,4)\mid K_n(2,3))$ between two
neighbor keystreams $K_n(2,4)$ and $K_n(2,3)$, with $H(K_n(2,4))$ and $%
H(K_n(2,4)\mid K_n(2,3))$ being information entropy and conditional entropy,
respectively. The behaviors of Fig.2 are not changed if we take any other
pais of sites. Thus, all the keystreams produced by different sites are
practically independent from each other, and can be satisfactorily used for
the parallel encryption (decryption) purpose.

Second, we evaluate the practical security of the cryptosystem, i.e., the
resistance of the chaotic system against the plaintext-known and
public-structure attacks, by applying the error function analysis (EFA)\cite
{14}. The error function $e({\bf b})$ can be computed with the available
information and an arbitrary test key ${\bf b}$ as 
\begin{eqnarray}
e_{j_1,j_2}({\bf b}) &=&\frac 1T\sum\limits_{n=1}^T\left| i_n^{^{\prime
}}(j_1,j_2)-i_n(j_1,j_2)\right| ,\text{ \ \ }  \eqnum{4} \\
\text{\ }i_n(j_1,j_2) &=&\frac{I_n(j_1,j_2)}{2^{32}},\text{ \ \ \ }%
i_n^{^{\prime }}(j_1,j_2)=\frac{I_n^{^{\prime }}(j_1,j_2)}{2^{32}}  \nonumber
\end{eqnarray}
By varying ${\bf b}$ any third party may find the basin of $e_{j_1,j_2}({\bf %
b})$ and locate the key position ${\bf b}={\bf a}$ by identifying the
minimum error $e({\bf b=a})=0$, and then illegally decode any future
plaintexts.

In Figs.3(a)-(c) we fix $b_2=b_3=3.9$ and plot $e(b_1)$ vs $b_1$. $e(b_1)$
has a basin around its minimum $e(b_1=a_1=3.9)=0$, and the basin is
extremely narrow, showing the high sensitivity of $e(b_1)$ to key $b_1$.
Away from the basin $e(b_1)$ has a flat distribution fluctuating around its
average value. In Fig.3(d) we plot $e(b_1,b_2)$ in $b_1$-$b_{2\text{ }}$%
plane with $b_3=a_3=3.9$, an extremely small $e(b_1,b_2)$ basin hole is
located in the 2D parameter space. The $a_{j\text{ }}$parameter region
available for spatiotemporal chaos is at least in $a_j=[3.6,4.0]$, $j=1,2,3$%
. By applying the analysis similar to\cite{14} the volume of the key basin
can be estimated from Fig.3 as $V<(10^{-12})^3$ $\approx 10^{-36}$, and the
probability to find the key basin by an arbitrary test is $P<0.4^{-3}\times
10^{-36}\approx 10^{-35}$. From the flat distribution of $e(b)$ in
Figs.3(a), (b), (d) and from the numerous local minima of $e(b)$ in Fig.3(c)
one can hardly find any adaptive approach to reveal the $e(b)$ basin more
effective than random tests, and the third party has to make at least $%
10^{35}$ tests to find the key basin. This can be done by the currently best
computer in the world at least for $10^{16}$ years (the estimation can be
seen in\cite{14}).

Apart from the EFA method, there are many other known attack methods in both
conventional and chaotic cryptographies, such as linear and differential
attacks \cite{16}, nonlinear dynamic forecasting attacks\cite{10,11}, and so
on. All these methods are based on predicting plaintexts by revealing the
plaintext-induced statistical changes of ciphertexts. We have computed
probability distributions of the ciphertexts generated at different sites
for various plaintexts. It is clearly observed that for essentially
different plaintexts the probabilities of ciphertexts have the same uniform
distributions, and they are not distinguishable from each other. Therefore,
the above attacks can be hardly practical. Analytic inverse computation can
also break the security of Eq.(1). A simple evaluation shows that for our
system this inverse computation requires cost much more than the EFA method.

Our cryptosystem is better than AES with security in the following two
aspects. First, with secure key well hidden our system is practically
one-time pad cipher, while AES is definitely not either with hidden key.
Second, our system can easily increase its security level, e.g., increasing $%
m$ by one we can increase the security factor by $10^{12}$ with almost no
encryption speed reduction. For AES there exists a maximum security level ($%
2^{128}$ for given plaintext attack). For further increasing security from
this maximum level one has to increase the block length, and then
considerably decrease the encryption speed. Therefore, our system can much
easier in resistance against stronger attacks developed by new powerful
computers, like possible quantum computers.

The last point to be emphasized is that the cryptographic structure of
Eq.(1) makes the secure communication robust against channel noise
disturbances. All self-synchronizing cryptosystems have a disadvantage of
error avalanche, i.e., one bit error in ciphertext may cause a large number
of error bits in the received plaintext due to the finite synchronization
recovering time, and slower synchronization may cause more serious avalanche
effect. In this regard, our system has several useful advantages. First, our
decoding system takes a strong coupling $(1-\varepsilon )=10^{-2}<<1$ which
yields rather small largest Lyapunov exponent, leading to quickly damping of
any desynchronous disturbances in the receiver. This guarantees short
synchronization time (26 iterations for our parameter combination) and
relatively small error avalanche ($17$ iterations). Second, chaos
synchronization between the transmitter and the receiver can be realized by
using a single driving sequence. Thus, among all $M=25$ transmitted
ciphertext sequences, only one sequence is used for driving, i.e., only $%
\frac 1{25}$ transmitted bits has avalanche problem, and other bit errors do
not cause avalanche effect. Therefore, in average the avalanche destruction
can be considerably reduced. Third and the most important, in order to
reduce error avalanche people commonly include some additional bits for
protection of driving signal, that increases the costs of both cryptography
and signal transmission. In doing so our system has a great advantage of low
cost over AES because in the latter case one should protect all transmitted
bits (each has equal avalanche effect) while for the former only the driving
signal, i.e., $\frac 1M$ of the total transmitted bits, has the avalanche
effect and needs to be particularly protected.

The cryptosystem of Eqs.(1) and (2) with the given parameters has been
realized in a software experiment for duplex voice transmission by using
local university campus network. The experimental set is given by Fig.2 of%
\cite{14} with cryptosystem replaced by Eqs.(1) and (2) of this paper.
Experimental dialogue can be performed stably between two phones with
standard voice quality and standard speaking speed for arbitrarily long
time. Moreover, we have realized also experimental duplex voice transmission
by using normal city phone line where channel noise is considerably larger
than that of network. In this experiment our system works perfectly well
with high security ( high sensitivity to the key parameter change) and
satisfactory robustness against noise. A detailed discussion on this
experimental set and a detailed experimental comparison between our system
and AES will be reported soon.

This work was supported by the National Natural Science Foundation of China
under 10175010 and Nonlinear Science Project.

\strut

\newpage Captions of Figures

Fig.1 A schematic figure of the cryptosystem Eqs.(1) for $N=6,m=3$. All the $%
M=25$ square sites produce ciphers each 32 bits for an iteration while the
other six triangle sites not. All the driving and nondriving ciphers are
transmitted in a same open channel.

Fig.2(a) and (b) The mutual correlation $C_{24,23}(\tau )$ [Eq.(3.2) in [4]]
and the mutual information $I(K_n(2,4);K_n(2,3))$ between $K_n(2,4)$ and $%
K_n(2,3)$ vs time distance $\tau $ and sample number $G$, respectively. The
behaviors are not changed if we take any other pairs of keystreams.

Fig.3 (a)-(c) Error function $e(b_1)$ given in Eq.(4) plotted vs $b_1$, $%
b_2=b_3=3.9$. (d) $e(b_1,b_2)$ plotted in $b_1$--$b_2$ plane, $b_3=3.9$.

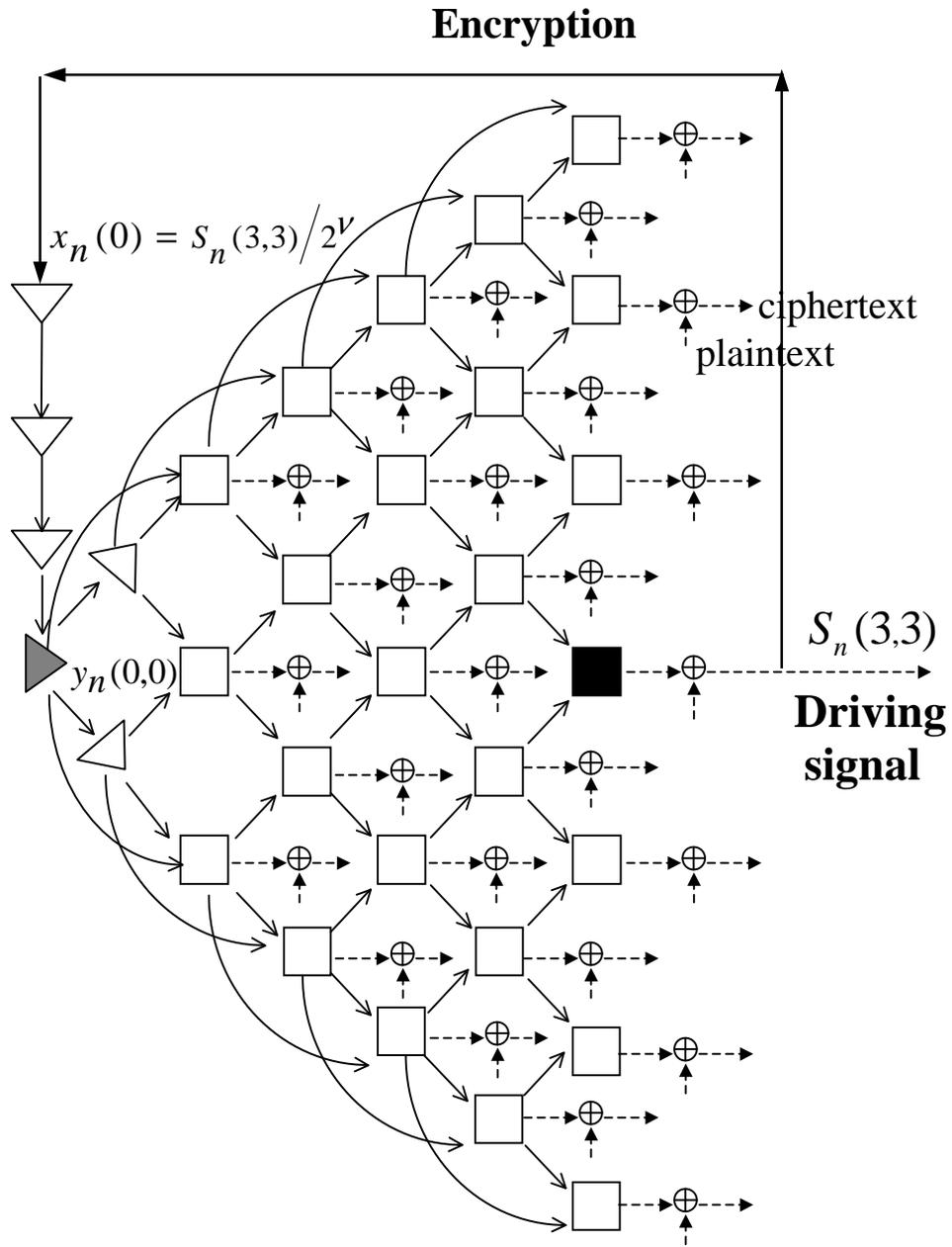
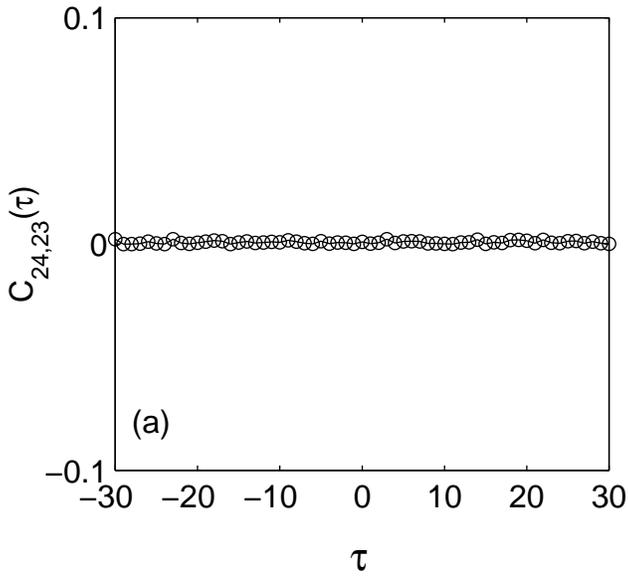
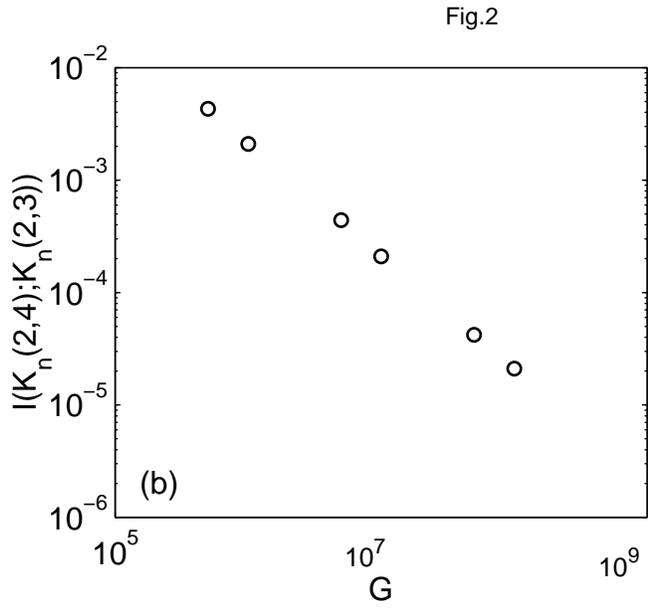
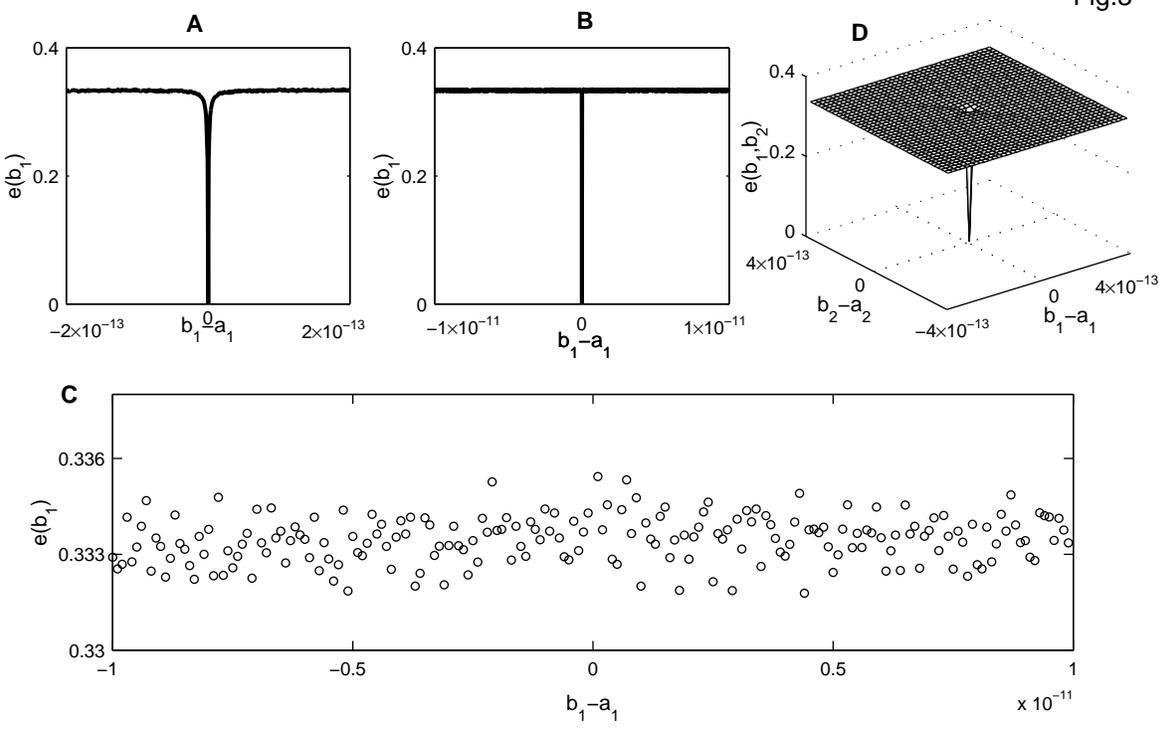
\end{document}